\documentstyle[12pt]{article}
\newcommand{\bibi}{\bibitem}
\newcommand{\etal}{\it {et al.}}
\newcommand{\prl}{\it Phys. Rev. Lett.}

\newcommand{\prb}{\it Phys. Rev. B}

\newcommand{\half}{\frac {1}{2}}
\newcommand{\third}{\frac {1}{3}}
\newcommand{\beq}{\begin{equation}}
\newcommand{\eeq}{\end{equation}\noindent}

\newcommand{\beqr}{\begin{eqnarray}}
\newcommand{\eeqr}{\end{eqnarray}\noindent}

\begin{document}
\title {Fermi Liquid - Non-Fermi
Liquid Transition in the Double Exchange Model}

\author{Sanjoy K. Sarker \\
Department of Physics and Astronomy \\
The University of Alabama, Tuscaloosa, Alabama 35487 \\
and\\
Joseph Henry Laboratories of Physics\\
Princeton University, Princeton, New Jersey 08544\\}
\date{}
\maketitle
\begin{abstract}

An approximate solution to the double-exchange model is presented.
The forced alignment of the conduction-electron spin with the core spin
that causes ferromagnetism also removes a large part of the
Hilbert space,
needed for coherent propagation of electrons carrying spin and charge.
As a result, the electron becomes a composite particle
and its Green's function exhibits a two-fluid character: a coherent
Fermi-liquid component associated with the ferromagnetically ordered core
spins and non-Fermi liquid component associated with the disordered spins.
With increasing temperature there is a continuous transfer of spectral
weight from the former to the latter until the Fermi liquid component
disappears above the ferromagnetic $T_c$. Implications for manganites,
which exhibit very large magnetorsistance, are discussed.

\end{abstract}

\vspace{0.1in}

\pagebreak

Recent discovery of very large magnetoresistance in
La$_{1-x}$Ca$_x$MnO$_3$ and other manganites \cite{jin,helm,chah}
has revived interest in the double-exchange model \cite{zen}.
Approximately in the range $0.2 < x < 0.4$, the low-temperature state
is a ferromagnetic metal which undergoes a transition to a paramagnetic
state above a critical temperature $T_c$.
The important effects originate from the d orbitals of the manganese ions.
The $E_g$ orbitals form a conduction band containing $1-x$ electrons per site.
The electron strongly interacts ferromagnetically via the Hund's rule
mechanism with the S = 3/2 core spin formed from the three-fold degenerate
$T_g$ orbitals. As it hops from one atom to the next,
the spin of a conduction electron must remain parallel
to the direction of the core spins. The latter will then tend to line up
to facilititate coherent propagation,
giving rise to metallic ferromagnetism \cite{zen,and}.
What is new and surprising is that, associated with the magnetic transition,
there appears to be a metal-insulator transition in the manganites and,
that it is the high-temperature phase that is insulating.
The large increase in resistivity at $T_c$ and the eneormous
magnetoresistance are clearly linked to this transition,
imply extreme sensitivity to external fields. The aim of this work
is to see whether such a sensitivity is intrinsic to the double-exchange
model.

We consider the following Hamiltonian
\beq H = - \sum _{ij\sigma} t_{ij} c^{\dag}_{i\sigma}c_{j\sigma}
-  J_H \sum _i \vec S_i\cdot \vec s_i. \eeq
Here $c_{i\sigma}$ destroys an $E_g$ electron at site i and the first
term is then the usual nearest-neighbor hopping Hamiltonian with
$t_{ij} = t$. The second term describes the Hund's rule coupling
between the core spin $\vec S_i$ and the conduction electron spin
density $\vec s_i$. We will focus on the $J_H \rightarrow \infty$
limit. The nominal two-fold degeneracy of the $E_g$ orbitals
could in fact be lifted by a coupling to the lattice.
For the most part we ignore the degeneracy,
its effect would be discussed later.

The key point is that the very cause of ferromagnetism, namely, that
the conduction-electron spin is forced to
follow the core-spin direction, removes a large portion of the Hilbert
space needed for coherent propagation of electrons carrying spin and charge.
Consider the on-site problem, with a classical core spin
$\vec S_i$. There are two single-electron states with electron
spin parallel and antiparallel to $\vec S_i$ and energies $\pm \half J_HS$,
respectively. As $J_H \rightarrow \infty$, both the doubly occupied and the
antiparallel singly occupied states are projected out.  In the
quantum mechanical case, an electron
combines with the core spin to form two manifolds of total spin
$S \pm \half$ with energies $- \half J_HS$ and $+ \half J_H(S+1)$,
respectively.
Again the $S - \half$ and the doubly occupied sectors are projected out.
This restriction is more stringent than that in the infinite-$U$ Hubbard model
where only double occupancy is forbidden, and dominates the physics.

We present approximate solutions that incorporate the projective physics
both below and above the transition by using a functional integral
representation similar to the one used by Millis {\etal} \cite{mil}.
We find that the conduction-electron Green's function is a sum of two terms:
a Fermi-liquid (coherent) component associated with the ferromagnetically
ordered core spin configuration, and a non-Fermi liquid (incoherent)
component associated with the disordered spin configurations.
With increasing temeperature ($T$), there is a continuous transfer of
spectral weight from the coherent to the incoherent component, as the
Fermi liquid component disappears completely above $T_c$. For $x < 0.5$, the
theory yields reasonable values for $T_c$ and a hole-like Fermi surface.
It is also found that the approximate particle-hole symmetry close to
the Fermi surface is violated.

Consider first the on-site problem with a classical $\vec S_i$,
pointing along some arbitrary direction.
Let $f_i$ and $g_i$ be operators that destroy single electrons with spins
parallel and antiparallel to $\vec S_i$.
The operators $c_{i\sigma}$ in
Eq. (1) are quantized along the global z-axis and can be represented as
a linear combination: $c_{i\sigma} = \frac{1}{\sqrt 2S}
 \lbrack b_{i\sigma}f_i + sgn(\sigma)b^*_{i,-\sigma}g_i\rbrack$,
where $b_{i\sigma} =  r_{i\sigma}e^{\phi_{i\sigma}}$ are complex numbers
which satisfy the normalization condition
$\sum _{\sigma}  b^*_{i\sigma}b_{i\sigma} = 2S$. If
$\vec S_i$ is expressed as $(S,\theta,\phi)$ in spherical coordinates,
then we
can choose $r_{i\uparrow} = \sqrt {2S} \cos {\frac{\theta _i}{2}}$,
{}~ $r_{i\downarrow} = \sqrt{2S} \sin {\frac{\theta _i}{2}}$ and
$\phi = \phi _{\downarrow} - \phi _{\uparrow}$. In the quantum case,
$b_{i\sigma}$ become time-dependent Bose fields, then it is convenient
to use functional integral techniques. Note that the core spin variables
can be expressed as: $S^+_i = b^*_{i\uparrow}b_{i\downarrow}$,
$S^z_i = \frac{1}{2} (b^*_{i\uparrow}b_{i\uparrow} -
b^*_{i\downarrow}b_{i\downarrow})$. This is nothing but the
exact Schwinger boson representation of the core spin \cite{aro,sar8}
-- a fact that will be useful later.
As $J_H \rightarrow \infty$, we can ignore the upper (antiparallel)
level (i.e. terms containing $g$) so that the conduction-electron fields
are exactly represented by
\beq c_{i\sigma} = \frac{1}{\sqrt {2S}} b_{i\sigma}f_i,\eeq
where $c_{i\sigma},f$ are Grassman field. The partition function can be
written as a functional integral with an action given by
\beq {\cal A} = \int_{0}^{\beta} d\tau \lbrack\sum _{i\sigma}(1 +
\frac{f^*_if_i}{2S})
b^*_{i\sigma}\frac {\partial b_{i\sigma}}{\partial\tau} +
(f^*_i\frac{\partial f_i}{\partial \tau} + \mu f^*_if_i) +  \frac{1}{2S}
\sum _{ij\sigma}t_{ij}b^*_{i\sigma}b_{j\sigma}f^*_if_j\rbrack. \eeq
The integrals over the Bose fields are to be done subject to
constraints $\sum _{\sigma}b^*_{i\sigma}b_{i\sigma} = 2S$.
The action is invariant under the gauge transformation:
$f_j \rightarrow f_je^{i\psi _j};~
b_{j\sigma} \rightarrow b_{j\sigma}e^{-i\psi _j}$. Hence at each site,
one of the Bose fields, say $b_{j\uparrow}$, can be chosen to be real. Then,
written in terms of the polar variables $(S,\theta,\phi)$,
 the action becomes identical to the one derived by Millis {\etal} \cite{mil}.

It is clear from Eqs. (2) and (3) that,
as a consequence of projecting out the antiparallel state,
the conduction electron, in essence, has become a composite object.
Its charge is carried by a {\em spinless}
fermion field $f$, and its spin is wedded to the core spin and is carried
by the Bose fields.  Thus, the charge of the conduction electron
is given by $n_i = f^*_if_i$ and its spin by $\vec s_i =
\frac{f^*_if_i}{2S}\vec S_i$. The last term in the action describes hopping
of a spinless charge characterized by a fluctuating hopping parameter
$tB_{ij}/S_R$,
where $B_{ij} \equiv \half \sum _{\sigma}b^*_{i\sigma}b_{i\sigma}$ essentially
measures the nearest-neighbor ferromagnetic correlation and has a magnitude
$|B_{ij}| = S\lbrack \cos ^2 (\frac{\theta _i - \theta _j}{2}) -
\sin \theta _i\sin \theta _j\cos ^2(\frac{\phi _i - \phi _j}{2})\rbrack. $
This has a maximum value (= $S$) for ferromagnetic alignment: $\theta _i =
\theta _j; \phi _i = \phi _j$. Eqs. (2) and (3) are very similar
with to the corresponding ones for the infinite-$U$
Hubbard model (or, $t-J$ model with $J = 0$). Physics is quite different,
however \cite{jay3}.  The charge and spin-fields in the present case are
not connected by
a projective constraint: \lq\lq spinons" and \lq\lq holons" can occupy the
same site simultaneously. The absence of such
gauge fluctuations is probably the reason why a stable ferromagnetic
state can exist in these systems for such large $x$.

A Hartree-Fock decomposition of (3) yields
effective actions that are quadratic in the charge and
spin-fields: ${\cal A}_{MF} = {\cal A}_f + {\cal A}_b + const$.
The factor $(1 + \frac{f^*_if_i}{2S})$ can be absorbed by
shifting $\mu$ and rescaling
the Bose fields so that they correspond to a spin of
renormalized magnitude: $S \rightarrow S_R = S\zeta = S +
\frac{1 -x}{2}$. This has a very simple meaning: the spin of a given
site fluctuates between $S$ (no electron) and $S + \half$ (one electron),
the average value for $1-x$ electrons is $S_R$.
The actions are then given by
\beq {\cal A}_b = \int _{0}^{\beta} d\tau \lbrack \sum _{i\sigma}
b^*_{i\sigma}\frac{\partial b_{i\sigma}}{\partial \tau} + \frac{D}{2S_R}
\sum _{ij\sigma} t_{ij}
b^*_{i\sigma}b_{j\sigma} \rbrack,  \eeq
\beq {\cal A}_f = \int _{0}^{\beta} d\tau \lbrack \sum _i
(f^*_i\frac{\partial f_i}{\partial \tau} - \mu f^*_if_i)
+ \frac{B}{S_R} \sum _{ij\sigma} t_{ij}f^*_if_j \rbrack,  \eeq
where $D = \langle f^*_if_j\rangle$ is the average \lq\lq kinetic energy" of
charge and $B = <B_{ij}>$ is the average value of
short-range ferromagnetic correlation. Thus,
although the bare problem has a single energy scale (the bare bandwidth
$W = 12t$), the propagation of charge and spin is governed by distinct
energy scales as characterized by the hopping parameters $t_f =
\frac{tB}{S_R}$ and $t_b = \frac{tD}{2S_R}$, respectively.
A determination of these energy scales is very important from an experimental
standpoint since $W$ can be in the eV range,
the experimental $T_c$ is only a few hundred degrees $\lbrack 1-3\rbrack$.

A number of results can be obtained independently of the constraints.
Since fermion fields are not constrained, ${\cal A}_f$ simply describes
free spinless fermions characterized by a
cosine band $\epsilon _f(k) = - 2t_f(\cos k_x + \cos k_y + \cos k_z)$.
The parameter $B$ is maximum (= $S_R$) in the ferromagnetic state
and decreases with increasing $T$, but must remain finite at $T_c$ since
it measures only {\em short-range} magnetic correlations. Hence, we
come to the conclusions that (1) charge fermions remain
\lq\lq metallic" across the feromagnetic transition and (2) the
charge bandwidth, $W_f = \frac{B}{S_R}W$, equals $W$
at $T = 0$, but decreases with increasing $T$ as $B$ decreases.
For fixed $S_R$, there is an obvious
symmetry about $x = 0.5$ (quarter filling), so we can restrict our
attention to $x < 0.5$. (3) The theory predicts a
hole-like Fermi surface in this region.  To determine $T_c$, we need
to compute $D$ which determines the Bose parameter $t_b$.
For $x$ not too large, we use a quadratic approximation for the hole spectrum
to obtain
$$ D =  x - \frac{x(6\pi ^2x)^{2/3}}{10}\lbrace 1 +
\frac{5\pi ^2}{12}(\frac{kT}{t_f\epsilon _F})^2\rbrace,$$
where $\epsilon _F = (6\pi ^2x)^{2/3}$ is the hole Fermi energy for
a band with $t_f = 1$. Thus $D$ is quite small, depends only on
$x$ at $T = 0$.  For  $x = 0.2$, $D \sim 0.1$ at
$T = 0$, and decreases with increasing $T$. Therefore,
$t_b/t_f = D/(2B) \sim D/(2S_R) << 1$. Hence, $T_c$ must be much smaller
than the fermion bandwidth.

The Bose Hartree-Fock problem
is identical to that for a spin-$S_R$ Heisenberg ferromagnet, but with a
{\em temperature dependent} exchange constant $J = tD/(2BS_R)$. For the
Heisenberg model, a qualitatively correct solution is obtained both
above \cite{aro}
and below \cite{sar8} the transition by imposing the
constraint on the average via a Lagrange multiplier $\Lambda$.
Then the bosons are
free, with a spectrum $\omega _k = 6t_b\gamma _k$, with $\gamma _k =
1 - \third(\cos k_x + \cos k_y + \cos k_z)$. Long-range magnetic order
(along the x direction)
appears through a Bose condensation in the $\vec k = 0$ mode \cite{sar8}.
The magnetization is the same as the condensate density and is given
by.
\beq m = S_R - \int \frac{d^3k}{(2\pi )^3}\frac{1}{e^{\beta (\omega _k
+ \Lambda)} - 1}\eeq
To obtain $B$ we only need to replace $1$ by $\gamma _k$ in the numerator.
The equations have been discussed in detail in ref.\cite{sar8}.  Briefly,
in the ordered regime, $\Lambda = 0$ and $m > 0$. The ground-state
is ferromagnetic since at $T = 0$ both
 integrals vanish giving $m = S_R$ and $B = S_R$. By expanding the
Bose spectrum we obtain the spin-wave theory results at low
temperatures: $m = S_R - const.~ T^{\frac{3}{2}}$ and $B = S_R -
const.~ T^{\frac{5}{2}}$. Above $T_c$, $m = 0$, and $\Lambda >
\propto \xi ^{-2} > 0$, where $xi$ is the (correlation) length over
which spins are ordered.

We have numerically and accurately
solved the fermion-boson self-consistency problem (for the exact
cosine spectra).  Fig 1 shows the critical temperature, $2kT_c/W$, as a
function of doping. Also shown are the
 chemical potential $2\mu_f/W$ and the charge fermion bandwidth $W_f/W =
t_f/t$, evaluated at $T_c$. Two aspects need to be
stressed. (1) $T_c$ is smaller than the fermion bandwidth by an
order of magnitude, confirming the earlier statement that the charge
degrees of freedom remain metallic across the ferromagnetic transition.
(2) The magnitude of $T_c$ is quite reasonable. Thus for a bare bandwidth
of 2 eV, $T_c$ ranges from 20 to 40 meV for 0.14 $< x < $ 0.5. Of course,
fluctuations would bring down $T_c$ somewhat.

If we take a two-fold degenerate level, there are two spinless charge
fermions at each site, otherwise the calculation is very similar.
This leads to a larger $T_c$ since the parameter
$D \sim (1-x)$ rather than $x$. It also implies an electron-like
Fermi surface since there are $\frac{(1-x)}{2}$ electrons per orbital.
However, the high-field Hall effect seems to be consistent with a
hole-like Fermi surface \cite{ong}. An experimental determination of the
Fermi surface would settle this issue.

{\em Spectral Function:} To gain further insight we
have calculated the electron Green's function. The electron
is a composite object carrying both spin and charge, and is therefore
a superposition of essentially infinite pairs of charge fermions
and spin bosons subject only to momentum and energy conservation:
$c_{k\omega\sigma} = (2S_RN\beta)^{-\half}
 \sum _{q\nu}b_{q\nu\sigma}f_{k-q,\omega - \nu}$, where $\omega$ and
$\nu$ are the odd and even Matsubara frequencies. Then, the
electron Green's function is a convolution of the fermion
and the Boson Green's functions \cite{sar9}. We do a particle-hole
transformation on the charge fermions ($h = f^*$) and work with holons,
characterized by a spectrum $\epsilon _k = -6t_f\gamma _k$.
The spectral function is given by
\beq A(k\omega) = \frac{m\pi}{S_R} \delta (\omega + \epsilon _k - \mu)
+ \frac{1}{8\pi ^2S_R} \int d^3q
\lbrack f(\epsilon _{q - k} - \mu) + n(\omega _q + \Lambda)\rbrack
\delta (\omega - \omega _q - \Lambda + \epsilon _{q-k} - \mu).\eeq
where $n(x)$ and $f(x)$ are Bose and Fermi functions.
Each term in the integral (sum) describes the
propagation of a charge riding on a particular mode of the
core spins. As is usual in Bose systems, the bosons
separate into a condensate or spin-ordered and
an excited or spin-disordered component. The condensate component (first
term) clearly constitutes a coherent Fermi liquid.
However, this is no ordinary Fermi liquid,
since it represents spinless fermions with a
spectral weight proportional to the {\em magnetization m},
and thus weakens with increasing $T$
and disappears at $T_c$, as the spectral weight is
continuously transferred to the second term which we call $A_1(k,\omega)$.
We now show that $A_1$ is incoherent, i.e., it does not have a
Fermi-liquid character and has a number unusual properties.

In the ordered state $\Lambda = 0$. Since
$\omega _q > 0$, it is is easy to see that, at $T = 0$,
 $A_1$ is nonzero only for $\omega > 0$. This reflects the
fact that we can only create (not destroy) an excited boson
at $T = 0$. Therefore, the approximate particle-hole symmetry near the Fermi
surface, a property of ordinary Fermi liquids, is violated.
This is a consequence of broken time-reversal invariance
and should be observable, perhaps in photoemission and tunneling
(density of states)
measurements. The density of states is easier to anlayze
and is given by
\beq D(\omega) = \frac{m}{2S_R}D_h(\mu - \omega) + \frac{1}{2S_R}
\int dz D_b(z)D_h(z + \Lambda + \mu - \omega)\lbrack n(z +
\Lambda) + f(z + \Lambda - \omega)\rbrack, \eeq
where $\omega $ is measured relative to the Fermi surface, and
$D_b$ and $D_h$ are the boson and fermion densities of states. The first
term is the condensate contribution. Since $D_b(z)$
nonzero only for $z > 0$, we see immediately that, at $T = 0$,
the integrand contributes only if $\omega > z > 0$, exhibiting the
particle-hole asymmetry. In the region $kT,\omega << \mu$, $D_h$ varies
slowly with frequency, and $D_b(\omega) = const.~~\Theta (\omega)\omega^{1/2}$
at low frequencies.  Then both above and below $T_c$ we obtain,
$D(\omega) = D_h(\mu + \Lambda - \omega)\lbrack \half +
\frac{1}{2S_R}\phi (\omega ,T)\rbrack$, where $\phi$ is the symmetry
violating contribution and has the scaling form
$\phi (\omega ,T) = \omega ^{3/2}g(\frac {\omega - \Lambda}{kT})$.
At $T = 0$, $\phi = \Theta (\omega) (\omega/\omega _m)^{3/2}$, where
$\omega _m$ is the maximum boson energy. For $T > 0$, $\phi $ has the
same $\omega ^{3/2}$ form for $\omega >> kT$, but acquires
a finite value for $\omega _ \le 0$, which vanishes exponentially as
$T^{3/2}e^{-\beta |\omega|}$ as $\omega \rightarrow -\infty$.
Precisely, on the Fermi surface ($\omega = 0$), $\phi $ is
temperature dependent and scales as $T^{3/2}$.

The spectral function itself can be calculated
 analytically by using the quadratic
approximations for the fermion and boson spectra:
$\epsilon _k \approx t_f(\vec Q - \vec k)^2,~~\omega _q \approx  t_bq^2$.
Let $\Omega \equiv
(1 - \frac{t_b}{t_f})(\omega - \Lambda+ \epsilon _k - \mu)/\epsilon _k$.
Then,
\beq A_1 = \frac{\Theta (1 - \Omega)\Theta
(\omega _m - \omega _-)}{16\pi St_ft_b\beta |Q - k|}
\lbrack \log \frac{1 + e^{-\beta(\omega_1 + \Lambda- \omega)}}
{1 + e^{-\beta(\omega _2 + \Lambda - \omega)}} +
\log \frac{1 - e^{-\beta(\omega_2 + \Lambda)}}
{1 - e^{-\beta(\omega _1 + \Lambda)}} \rbrack, \eeq
where $~~\omega _1 = \omega _-$,
$\omega _2 = min(\omega _m,\omega _+)$, $\omega _m$ is the maximum
boson energy and
$$\omega _{\pm} = \frac{2t_bt_f\epsilon _k}{(t_f - t_b)^2}
\lbrack 1 - \half \Omega \pm (1 - \Omega)^{\half} \rbrack. $$
Note that $\omega _+ > \omega _- \ge 0$. The quasiparticle peak,
if it exists, would occur at $\Omega = 0$.
At $T = 0$ only the first term in (9) contributes. It
has no singularity and exists only for $\omega > \omega _-
\ge 0$, exhibiting the lack of p-h symmetry. For $T > 0$,
this term just broadens out. The
second term in (9) contributes only for $T > 0$ and also has a
non-Fermi liquid character.  In the ordered region
($\Lambda = 0$) it exhibits a
 logarithmic singularity $\sim - \log \omega _-  \sim  - \log
\Omega ^2$. But as shown in Fig. 2, even this weak
singulrity disappears for $T > T_c$ i.e., for a finite $\Lambda$ as
small as $0.0025W$. Therefore, the state above $T_c$ is quite
remarkable: the charge of the electron is itinerant and has a Fermi surface,
but the spin of the electron is localized over a distance $\xi \sim
\Lambda ^{-\half}$. The electron itself does not exist as a well-defined
quasiparticle nor does it have a well defined Fermi-surface.

To summarize, the Green's function is a sum of two
terms: one representing a coherent, perfectly polarized, spinless
Fermi liquid, and the other an incoherent, non-Fermi liquid of
unpolarized electrons. With increasing $T$, there is a continuous
transfer of spectral weight from the coherent to the incoherent
component.  It is not hard to see that such a separation would also
occur for higher order Green's functions. This
is consistent with the continuous transfer of spectral weight in the
measured optical conductivity from a metallic to an incoherent
component \cite{oki}.  In this two-fluid picture,
 the dc conductivity is $\sigma = \sigma _c +
\sigma _{inc}$, where $\sigma _c = n_ce\mu _c$ is the coherent part, with
$n_c$ and $\mu _c$, the density and the
mobility of the \lq\lq coherent" carriers, respectively. There is of
course no transport theory for the incoherent electrons---their
interaction with the lattice \cite{hwa} and disorder may be very important.
But calculations based on scattering of itinerant electrons from spin
fluctuations is clearly not applicable. However, if we accept the
two-fluid picture, then, to be consistent with experiments,
the incoherent contribution $\sigma _{inc}$ is
similar to that for a semiconductor or Anderson-localized insulator. In
particular $\mu _{inc} << \mu _c$, perhaps by orders of magnitude.
Then, the resisitivity $\rho \sim
1/\sigma _c$, i.e., is metallic at low temperature. But close to $T_c$,
it rapidly goes over to $1/\sigma _{inc}$, simply because $n_c$ vanishes
as $T \rightarrow T_c$. The changing $n_c$ also allows $\rho$ to
change continuously without violating unitarity.  An applied magnetic
field has the opposite effect: it
transfers electrons from the incoherent component to the metallic component
by forcing the core spins to line up, the effect being strongest close to
$T_c$. In the insulating regime, introduction
of even a small number of coherent electrons would cause a large drop in the
resistivity {\em simply because of the very large difference in the mobility}.
Therefore, a modest field will result in large magnetoresistance as
well as a large shift in the postion of the resistivity maximum, as
observed.  This sensitivity to external fields is due to spin-charge
separation.

The author is grateful to C. Jayaprakash, T. L. Ho, P. W. Anderson and N. P.
Ong many useful discussions.  Work at Princeton was supported by a grant
from NSF (DMR 9104873).

\pagebreak

\centerline{Figure Captions}

Fig. 1. Energy Scales: Plot of $\frac{2kT_c}{W}$, $\frac{2\mu _f}{W}$ and
the $W_f/W$ as a function of doping $x$. Here $W = 12t =$ bare bandwidth,
$W_f = 12 t_f = $ Fermion bandwidth at $T_c$ and $\mu _f = $ is the hole
Fermi energy.

Fig. 2. The incoherent part of the spectral function $A_1$ close to
the Fermi surface. Below the
transition (solid line) it has a weak logarithmic singularity which
disappears above $T_c$ for $\Lambda = 0.0025 W$ (dotted line). The
parameter $G \equiv 2\Lambda/W$. Energies are in units of $W/2$.

\end{document}